%% file: eacl2021.tex
\documentclass[11pt,a4paper]{article}
\usepackage[hyperref]{eacl2021}
\usepackage{times}
\usepackage{latexsym}

\usepackage{makecell}
\usepackage{pgfplots}
\usepackage{xcolor}
\usepackage{amsmath,amsfonts,amssymb}
\usepackage[multiple]{footmisc}

\usepackage{multirow}
\usepackage{tabularx}
\usepackage{arydshln}

\usepackage{microtype}

\aclfinalcopy 

\usepackage{xr}
\makeatletter
\newcommand*{\addFileDependency}[1]{
  \typeout{(#1)}
  \@addtofilelist{#1}
  \IfFileExists{#1}{}{\typeout{No file #1.}}
}
\makeatother

\newcommand{\bnh}[1]{\underline{#1}}
\newcommand{\bs}[1]{\textbf{#1}}

\title{Zero-shot Neural Passage Retrieval via\\Domain-targeted Synthetic Question Generation}

\author{
Ji Ma\quad Ivan Korotkov\quad Yinfei Yang\quad Keith Hall \quad Ryan McDonald\\
Google Research\\
  \texttt{\{maji,ivankr,yinfeiy,kbhall,ryanmcd\}@google.com} \\}

\date{}

\begin{document}
\maketitle
\begin{abstract}
A major obstacle to the wide-spread adoption of neural retrieval models is that they require large supervised training sets to surpass traditional term-based techniques, which are constructed from raw corpora.
In this paper, we propose an approach to zero-shot learning for passage retrieval that uses synthetic question generation to close this gap. The question generation system is trained on general domain data, but is applied to documents in the targeted domain. This allows us to create arbitrarily large, yet noisy, question-passage relevance pairs that are domain specific. Furthermore, when this is coupled with a simple hybrid term-neural model, first-stage retrieval performance can be improved further. 
Empirically, we show that this is an effective strategy for building neural passage retrieval models in the absence of large training corpora. Depending on the domain, this technique can even approach the accuracy of supervised models.
\end{abstract}

\section{Introduction}

\label{sec:intro}

\begin{figure}
    \centering
    \includegraphics[width=1.0\columnwidth]{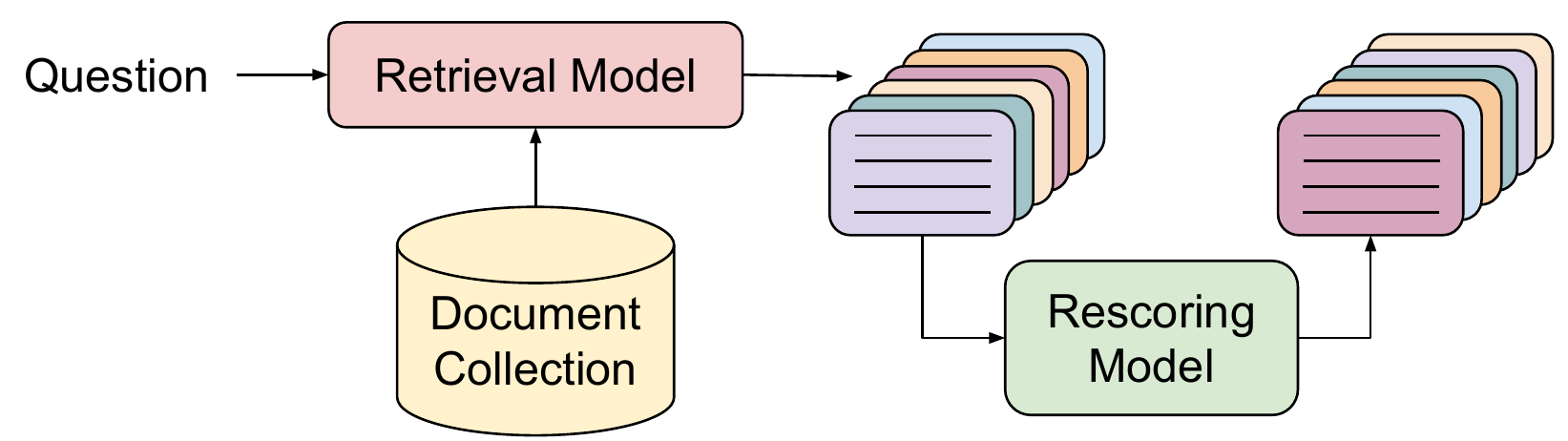}
    \vspace{-0.2in}
    \caption{End-to-end neural retrieval. A first-stage model over a large collection returns a smaller set of relevant passages which are reranked by a rescorer.}
    \label{fig:ret-rnk}
    \vspace{-0.1in}
\end{figure}

Recent advances in neural retrieval have led to advancements on several document, passage and knowledge-base benchmarks \cite{Guo2016ADR,pang2016text,hui2017pacrr,dai2018convolutional,GillickE2E,nogueira2019passage,macavaney2019cedr,yang2019end,yang2019simple,use-qa}.
Most neural passage retrieval systems are, in fact, two stages \cite{zamani2018neural,yilmaz2019cross}, illustrated in Figure~\ref{fig:ret-rnk}. The first is a true retrieval model (aka first-stage retrieval\footnote{Also called \emph{open domain} retrieval.}) that takes a question and retrieves a set of candidate passages from a large collection of documents. This stage itself is rarely a neural model and most commonly is an term-based retrieval model such as BM25 \cite{robertson2004simple,yang2017anserini}, though there is recent work on neural models \cite{zamani2018neural,dai2019context,chang-ict,dense-passage-qa,luan-sparse-dense}. This is usually due to the computational costs required to dynamically score large-scale collections. Another consideration is that BM25 is often high quality \cite{lin2019neural}. After first-stage retrieval, the second stage uses a neural model to rescore the filtered set of passages. Since the size of the filtered set is small, this is feasible. 

The focus of the present work is methods for building neural models for \emph{first-stage passage retrieval} for large collections of documents. While rescoring models are key components to any retrieval system, they are out of the scope of this study. Specifically, we study the \emph{zero-shot} setting where there is no target-domain supervised training data \cite{xian2018zero}. This is a common situation, examples of which include enterprise or personal search environments \cite{hawking2004challenges,chirita2005using}, but generally any specialized domain.

The zero-shot setting is challenging as the most effective neural models have a large number of parameters, which makes them prone to overfitting. Thus, a key factor in training high quality neural models is the availability of large training sets. To address this, we propose two techniques to improve neural retrieval models in the zero-shot setting.

First, we observe that general-domain question-passage pairs can be acquired from community platforms \cite{shah2010evaluating,duan2017question} or high quality academic datasets that are publicly available \cite{NQ,bajaj2016ms}. Such resources have been used to create open domain QA passage retrieval models. However, as shown in \citet{multireqa} and in our later experiments, neural retrieval models trained on the general domain data often do not transfer well, especially for specialized domains.

\begin{figure}
    \centering
    \includegraphics[width=0.8\columnwidth,height=6.2cm]{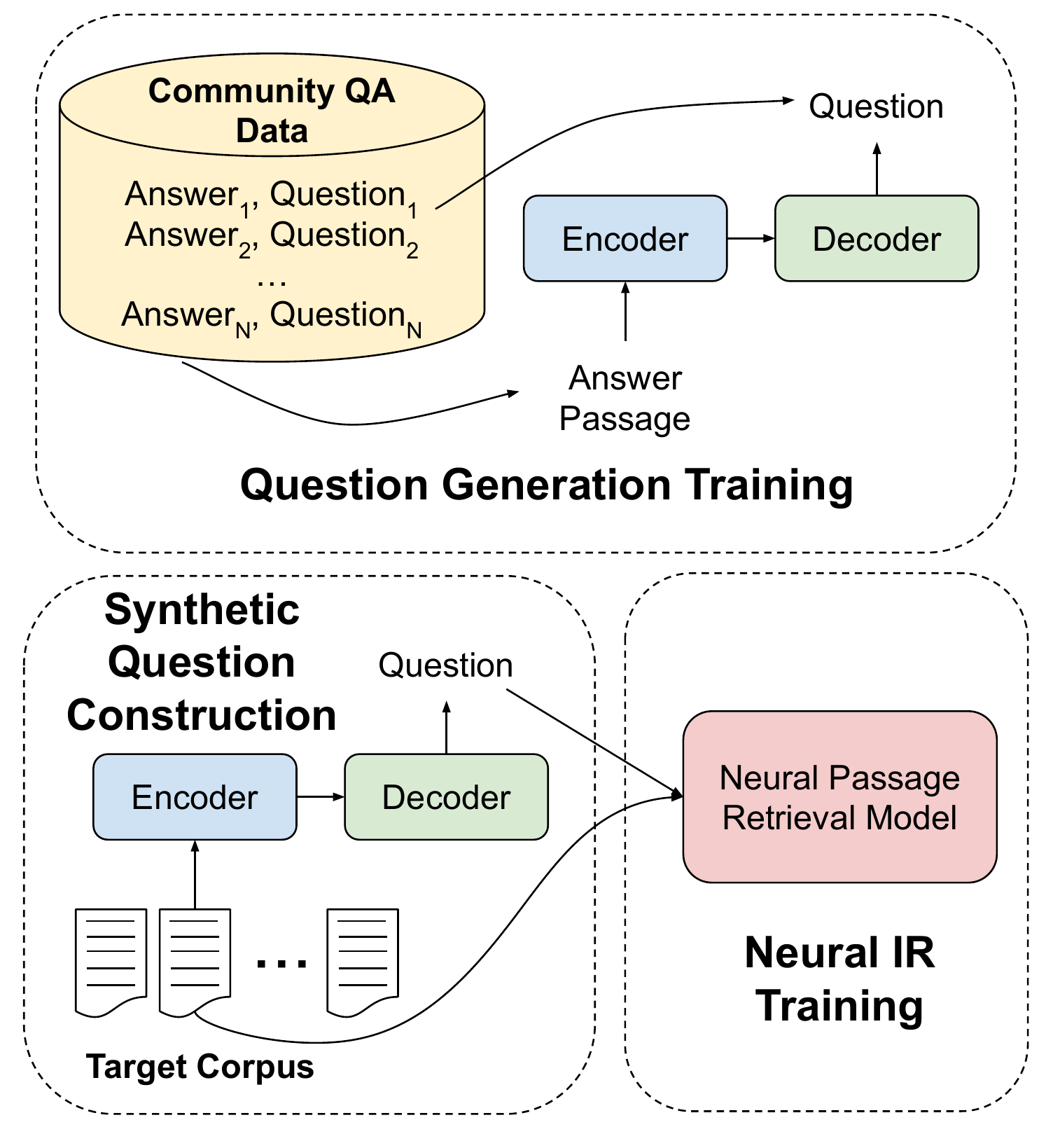}
    \vspace{-0.1in}
    \caption{Synthetic query generation for neural IR.}
    \label{fig:pipeline}
    \vspace{-0.15in}
\end{figure}

Towards zero-shot neural retrieval with improved domain adaptability,
we propose a data augmentation approach \cite{wong2016understanding} that leverages these naturally occurring question/answer pairs to train a generative model that synthesizes questions given a text \cite{zhou2017neural}.
We apply this model to passages in the target domain to generate unlimited pairs of synthetic questions and target-domain passages. This data can then be used for training. This technique is outlined in Figure~\ref{fig:pipeline}.

A second contribution is a simple hybrid model that interpolates a traditional term-based model -- BM25 \cite{robertson1995okapi} -- with our zero-shot neural model. BM25 is also zero-shot, as its parameters do not require supervised training. Instead of using inverted index which is commonly used in term-based search, we exploit the fact that BM25 and neural models can be cast as vector similarity (see Section~\ref{sec:hybrid}) and thus nearest neighbour search can be used for retrieval \cite{liu2011hashing,JDH17}. The hybrid model takes the advantage of both the term matching and semantic matching.

We compare a number of baselines including other data augmentation and domain transfer techniques. We show on three specialized domains (scientific literature, travel and tech forums) and one general domain that the question generation approach is effective, especially when considering the hybrid model. Finally, for passage retrieval in the scientific domain, we compare with a number of recent supervised models from the BioASQ challenge, including many with rescoring stages. Interestingly, the quality of the zero-shot hybrid model approaches supervised alternatives.

\section{Related Work}

\paragraph{Neural Retrieval} The retrieval vs.\ rescorer distinction (Figure~\ref{fig:ret-rnk}) often dictates modelling choices for each task. For first-stage retrieval, as mentioned earlier, term-based models that compile document collections into inverted indexes are most common since they allow for efficient lookup \cite{robertson2004simple,yang2017anserini}. However, there are studies that investigate neural first-stage retrieval. A common technique is to learn the term weights to be used in an inverted index \cite{zamani2018neural,dai2019context, dai_term_weighting2020}. Another technique is representation-based models that embed questions and passages into a common dense subspace \cite{palangi2016deep} and use nearest neighbour search for retrieval \cite{liu2011hashing,JDH17}. Recent work has shown this can be effective for passage scoring \cite{chang-ict,dense-passage-qa, MacAvaneyet2020}. 
Though all of the aforementioned first-stage neural models assume supervised data for fine-tuning. 
For rescoring, scoring a small set of passages permits computationally intense models. These are often called interaction-based, one-tower or cross-attention models and numerous techniques have been developed \cite{Guo2016ADR,hui2017pacrr,xiong2017end,dai2018convolutional,DeepRelevenceRanking}, many of which employ pre-trained contextualized models \cite{nogueira2019passage,macavaney2019cedr,yang2019end,yang2019simple}.
\newcite{Colbert} also showed that by delaying interaction to the last layer, one can build a first stage retrieval model which also leverages the modeling capacity of an interaction based models.

\paragraph{Model Transfer} Previous work has attempted to alleviate reliance on large supervised training sets by pre-training deep retrieval models on weakly supervised data such as click-logs \cite{neural_click,nir_weak_supervision}. Recently, \newcite{yilmaz2019cross} has shown that training models on general-domain corpora adapts well to new domains without targeted supervision. Another common technique for adaptation to specialized domains is to learn cross-domain representations  \cite{cross_domain_adv, DomainAdaEmail}.
Our work is more aligned with methods like \newcite{yilmaz2019cross} which use general domain resources to build neural models for new domains, though via different techniques -- data augmentation vs.\ model transfer. Our experiments show that data augmentation compares favorably a model transfer baseline. For specialized domains, recently, there have been a number of studies using cross-domain transfer and other techniques for biomedical passage retrieval via the TREC-COVID challenge\footnote{\texttt{ir.nist.gov/covidSubmit/}}\footnote{\texttt{ir.nist.gov/covidSubmit/archive.html}} that uses the CORD-19 collection \cite{wang2020cord}.

Question generation for data augmentation is a common tool, but has not been tested in the pure zero-shot setting nor for neural passage retrieval.
\newcite{duan2017question} use community QA as a data source, as we do, to train question generators. The generated question-passage pairs are not used to train a neural model, but QA is instead done via question-question similarity. Furthermore, they do not test on specialized domains. \newcite{alberti2019synthetic} show that augmenting supervised training resources with synthetic question-answer pairs can lead to improvements.  \newcite{nogueira2019document} employed query generation in the context of first-stage retrieval. In that study, the generated queries were used to augment documents to improve BM25 keyword search. Here we focus on using synthetic queries to train the neural retrieval models.

\paragraph{Hybrid Models} Combining neural and term-based models have been studied, most commonly via linearly interpolating scores in an approximate re-ranking stage \cite{dense-passage-qa,luan-sparse-dense} or through the final layer of a rescoring network \cite{severyn-etal-2015-distributional,DeepRelevenceRanking}. Since rescoring can be cast as classification, blending signals is straight-forward. However, this is approximate as it does not operate over the whole collection. For first-stage retrieval, the most common method is to learn term weights for a standard inverted index in order to make search efficient \cite{zamani2018neural,dai2019context}. Here we propose a first-stage retrieval model that incorporates both term-based (sparse) and neural-based (dense) representations in a hybrid model that uses nearest neighbor search for exact inference \cite{liu2011hashing,JDH17,wu2019efficient}. Similar methods using approximate nearest neighbour search have been investigated by \newcite{seo-etal-2019-real}.

\section{Synthetic Question Generation}
\label{sec:qgen}

\begin{table}
    \centering
    \scriptsize
    \begin{tabular}{p{2.8in}}
        \textbf{Ubuntu Forums}\\ \hline
        \textbf{Passage}: Every time I get a notification about and begin updating when they become available, the process is interrupted by an error message: error in foomatic-filters. Then I get ``error in linux generic package'' and a bunch of numbers. This is replaced before I can write it all down with ``error in Linux package''  Everything seems to go OK except I don't know if the updates are really being installed. I tried un-installing and re-installing foomatic-filters $\ldots$
        \\
        \textbf{Generated Question}: How do I get rid of error in foomatic-filters? \\
        \\ 
        \textbf{Biomedical Literature}\\ \hline
        \textbf{Passage}: Electroencephalographic tracings of 50 patients who presented the classical features of Friedreich's ataxia were reviewed $\ldots$
        Friedreich's ataxia is mainly a spinal disorder. Involvement of supraspinal and in particular brain stem or diencephalic structures may be more extensive in those patients who show electrographic abnormalities. This would require confirmation with comparative data based on pathological observations. Impaired function of brain stem inhibitory mechanism may be responsible for the slightly raised incidence of seizures in patients with Friedreich's ataxia and other cerebellar degenerations.
        \\
        \textbf{Generated Question}: What is the significance of Friedreich's ataxia? \\
    \end{tabular}
    \vspace{-0.05in}
    \caption{Examples of domain-targeted synthetic generated questions used to train passage retrieval models.}
    \label{tab:qgen_examples}
    \vspace{-0.1in}
\end{table}

In this work, we are specifically investigating the zero-shot scenario where there exists neither user issued questions nor domain specific data except the passage collection itself. We propose to address the training data scarcity issue by generating synthetic questions \cite{zhou2017neural,duan2017question,alberti2019synthetic,nogueira2019document}.
Leverage the fact that there are large question-answer data sources freely available from the web \cite{shah2010evaluating,duan2017question}. 
we first train a question generator using general domain question-answer pairs.
The passage collection of a target domain is then fed into this generator to create pairs of noisy question-passage pairs, which are used to train a retrieval model (see Figure~\ref{fig:pipeline}). 
In this work, we mine English question-answer pairs from community resources, primarily StackExchange\footnote{archive.org/details/stackexchange} and Yahoo! Answers\footnote{webscope.sandbox.yahoo.com/catalog.php?datatype=l}. 
Note we use stackexchange as it covers a wide range of topics, and we focus on investigating the 
domain adaptability of using a question generation approach.  
We leave comparing question generator trained on different datasets or using different architectures to future work.

To ensure data quality, we further filter the data by only keeping question-answer pairs that were positively rated by at least one user on these sites.
In total, the final dataset contains 2 millions pairs, and the average length of questions and answers are 12 tokens and 155 tokens respectively. This dataset is \textit{general domain} in that it contains question-answer pairs from a wide variety of topics.

Our question generator is an encoder-decoder with Transformer \cite{transformer} layers, which is a common  for generation tasks such as translation and summarization \cite{transformer,berts2s}. The encoder is trained to build a representation for a text and the decoder generates a question for which that text is a plausible answer. Appendix~\ref{app:qgen_details} has model specifics.

Our approach is robust to domain shift as the generator is trained to create questions based on a given text. As a result, generated questions stay close to the source passage material. Real examples are shown in Table~\ref{tab:qgen_examples} for technical and biomedical domains, highlighting the model's adaptability.

\section{Neural First-stage Retrieval}
\label{sec:neuralretrieval}

In this section we describe our architecture for training a first-stage neural passage retriever. Our retrieval model belongs to the family of \textit{relevance-based dense retrieval} \footnote{A.k.a.\ two-tower, dual encoder or dense retrieval.} that encodes pairs of items in dense subspaces \cite{palangi2016deep}.  
Let $Q = (q_1, \ldots q_n)$ and $P = (p_1, \ldots, p_m)$ be a question and passage of $n$ and $m$ tokens respectively.
Our model consists of two encoders, $\{f_Q(), f_P()\}$ and a similarity function, $\text{sim}()$.  
An encoder is a function $f$ that takes an item $x$ as input and outputs a real valued vector as the encoding, The similarity function, $\text{sim}()$, takes two encodings, $\textbf{q}, \textbf{p} \in \mathbb{R}^N$  and calculates a real valued score, $s = \text{sim}(\textbf{q}, \textbf{p})$.
For passage retrieval, the two encoders are responsible for computing dense vector representation of questions and passages.

\subsection{BERT-based Encoder}
\label{nn:enc}

In this work, both query and document encoders are based on BERT \cite{bert},
which has been shown to lead to large performance gains across a number of tasks, including document ranking~\cite{nogueira2019passage,macavaney2019cedr,yang2019simple}. 
In addition, we share parameters between the query and passage encoder -- i.e., $f_Q = f_P$, so called Siamese networks -- as we found this greatly increased performance while reducing parameters.

We encode $P$ as $(\text{CLS, } p_1, \ldots, p_m, \text{ SEP})$. For some datasets, a passage contains both a title $T=(t_1, ..., t_l)$ and content $C = (c_1, ..., c_o)$, in which case we encode the passage as $(\text{CLS, } t_1, ..., t_l, \text{SEP}, c_1, ..., c_o, \text{ SEP})$.
These sequences are fed to the BERT encoder. 
Let $h_{\text{CLS}} \in \mathbb{R}^N$ be the final representation of the ``CLS'' token.
Passage encodings $\textbf{p}$ are computed by applying a linear projection, i.e., $\textbf{p} = \textbf{W}*h_{\text{CLS}}$, where \textbf{W} is a $N\times N$ weight matrix (thus $N=768$), which preserves the original size of  $h_{\text{CLS}}$. This has been shown to perform better than down-projecting to a lower dimensional vector \cite{luan-sparse-dense}, especially for long passages.

We encode $Q$ as $(\text{CLS, } q_1, q_2, ..., q_n, \text{ SEP})$ which is then fed to the BERT encoder. 
Similarly, a linear projection on the corresponding ``CLS'' token, using the same weight matrix $\textbf{W}$, is applied to generate $\textbf{q}$.
Following previous work \cite{luan-sparse-dense, lee2019latent}, we use dot product as the similarity function, i.e., $ \text{sim}(\textbf{q}, \textbf{p}) = \langle \textbf{q}, \textbf{p} \rangle = \textbf{q}^\intercal \textbf{p}$.

\begin{figure}
    \centering
    \includegraphics[width=0.95\columnwidth]{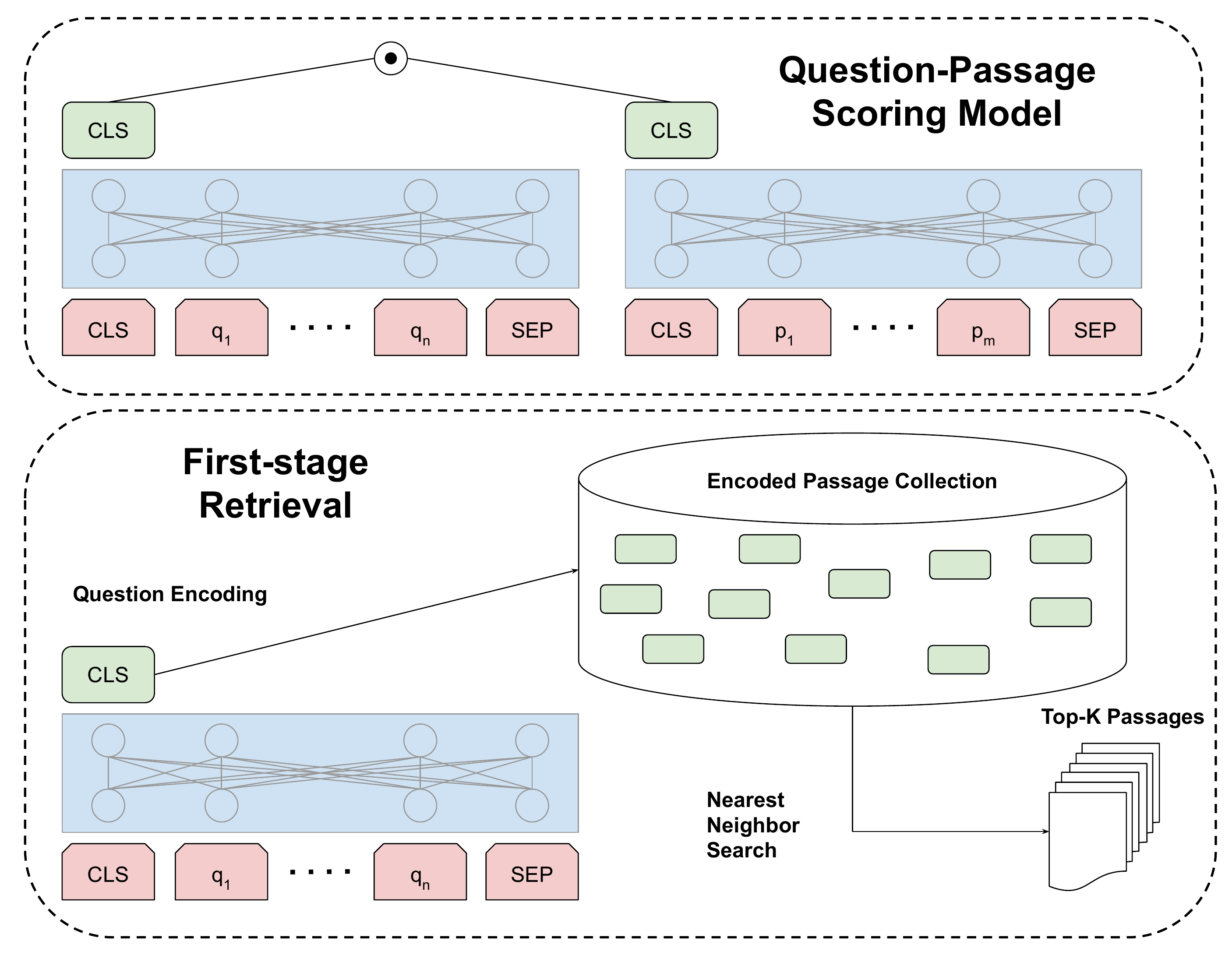}
    \vspace{-0.1in}
    \caption{First-stage neural passage retrieval. Top: A BERT-based transformer encodes questions and passages and scores them via dot-product. Bottom: Passages from the collection are encoded and stored in a nearest neighbour search backend. At inference, the question is encoded and relevant passages retrieved.}
    \label{fig:debert}
    \vspace{-0.1in}
\end{figure}

The top half of Figure~\ref{fig:debert} illustrates the model.

\subsection{Training}
For training, we adopt softmax cross-entropy loss.
Formally, given an instance $\{\textbf{q}, \textbf{p}^+, \textbf{p}_1^-, ..., \textbf{p}_k^-\}$ which comprises one query $\textbf{q}$, one relevant passage $\textbf{p}^+$ and $k$ non-relevant passages $\textbf{p}_i^-$.
The objective is to minimize the negative log-likelihood:
\begin{multline}
L(\textbf{q}, \textbf{p}^+, \textbf{p}_1^-, ..., \textbf{p}_k^-) = 
\\  
 \log(e^{\langle \textbf{q}, \textbf{q}^+ \rangle} + \sum_{i=1}^k{e^{\langle \textbf{q}, \textbf{q}_i^- \rangle}}) - \langle \textbf{q}, \textbf{q}^+ \rangle \nonumber
\end{multline}
This loss function is a special case of ListNet loss \cite{learning-to-rank-softmax} where all relevance judgements are binary, and only one passage is marked relevant for each training example.  

For the set $\{\textbf{p}_1^-, ..., \textbf{p}_k^-\}$, we use in-batch negatives.
Given a batch of (query, relevant-passage) pairs, negative passages for a query are passages from different pairs in the batch.
In-batch negatives has been widely adopted as it enables efficient training via  computation sharing \cite{yih-etal-2011-learning, GillickE2E, dense-passage-qa}.

\subsection{Inference}

Since the relevance-based model encodes questions and passages independently, we run the encoder over every passage in a collection offline to create a distributed lookup-table as a backend. At inference, we run the question encoder online and then perform nearest neighbor search to find relevant passages, as illustrated in the bottom half of Figure~\ref{fig:debert}.
While there has been extensive work in fast approximate nearest neighbour retrieval for dense representations \cite{liu2011hashing,JDH17}, we simply use distributed brute-force search as our passage collections are at most in the millions, resulting in exact retrieval.

\subsection{Hybrid First-stage Retrieval}
\label{sec:hybrid}

Traditional term-based methods like BM25 \cite{robertson1995okapi} are powerful zero-shot models and can outperform supervised neural models in many cases \cite{lin2019neural}. Rescoring systems have shown that integrating BM25 into a neural model improves performance \cite{DeepRelevenceRanking}. However, for first-stage retrieval most work focuses on approximations via re-ranking \cite{dense-passage-qa,luan-sparse-dense}. Here we present a technique for exact hybrid first-stage retrieval without the need for a re-ranking stage. Our method is motivated by the work of \newcite{seo-etal-2019-real} for sparse-dense QA.

For a query $Q$ and a passage $P$, BM25 is computed as the following similarity score, 
\begin{multline}
\text{BM25}(Q, P) = \\
\sum_{i=1}^n  \frac{\text{IDF}(q_i) * \text{cnt}(q_i \in P)*(k + 1)}{\text{cnt}(q_i \in P) + k*(1 - b + b * \frac{m}{m_\text{avg}})}, \nonumber
\end{multline}
where $k$/$b$ are BM25 hyperparameters, IDF is the term's inverse document frequency from the corpus, cnt is the term's frequency in a passage, $n$/$m$ are the number of tokens in $Q$/$P$, and $m_\text{avg}$ is the collection's average passage length.

Like most TF-IDF models, this can be written as a vector space model. Specifically, let $\textbf{q}^{\text{bm25}} \in [0,1]^{|V|}$ be a sparse binary encoding of a query of dimension $|V|$, where $V$ is the term vocabulary. Specifically this vector is 1 at position $i$ if $v_i \in Q$, here $v_i$ is the $i$-th entry in $V$.
Furthermore, let $\textbf{p}^{\text{bm25}} \in \mathbb{R}^{|V|}$ be a sparse real-valued vector where,
\begin{multline}
\textbf{p}_i^\text{bm25} = 
 \frac{\text{IDF}(v_i)* \text{cnt}(v_i \in P)*(k + 1)}{\text{cnt}(v_i \in P) + k*(1 - b + b * \frac{m}{m_\text{avg}})} \nonumber
\end{multline}
We can see that,
\begin{equation}
    \text{BM25}(Q, P) = \langle \textbf{q}^\text{bm25}, \textbf{p}^\text{bm25} \rangle \nonumber
\end{equation}
As BM25 score can be written as vector dot-product, this gives rise to a simple hybrid model,
\begin{align}
    \nonumber
    \text{sim}(\textbf{q}^\text{hyb}, \textbf{p}^\text{hyb}) &= \langle \textbf{q}^\text{hyb}, \textbf{p}^\text{hyb} \rangle \\ \nonumber
    &= \langle [\lambda\textbf{q}^\text{bm25},\textbf{q}^\text{nn}], [\textbf{p}^\text{bm25}, \textbf{p}^\text{nn}] \rangle \\ \nonumber
    &= \lambda \langle \textbf{q}^\text{bm25}, \textbf{p}^\text{bm25} \rangle + \langle \textbf{q}^\text{nn}, \textbf{p}^\text{nn} \rangle,
\end{align}
where $\textbf{q}^\text{hyb}$ and $\textbf{p}^\text{hyb}$ are the hybrid encodings that concatenate the BM25 ($\textbf{q}^\text{bm25}$/$\textbf{p}^\text{bm25}$) and the neural encodings ($\textbf{q}^\text{nn}$/$\textbf{p}^\text{nn}$, from Sec~\ref{nn:enc}); and $\lambda$ is a interpolation hyperparameter that trades-off the relative weight of BM25 versus neural models.

Thus, we can implement BM25 and our hybrid model as nearest neighbor search with hybrid sparse-dense vector dot-product \cite{wu2019efficient}.

\section{Experimental Setup}

We outline data and experimental details. The Appendix has further information to aid replicability.

\vspace{-1mm}
\subsection{Evaluation Datasets}
\label{sec:eval_data}
\paragraph{BioASQ} Biomedical questions from Task B Phase A of BioASQ \cite{BIOASQ}.  
We use BioASQ 7 and 8 test data for evaluation.
The collection contains all abstracts from MEDLINE articles.
Given an article, we split its abstract into chunks with sentence boundaries preserved.  
A passage is constructed by concatenating the title and one chunk.
Chunk size is set so that each passage has no more than 200 wordpiece tokens.

\vspace{-1mm}
\paragraph{Forum} Threads from two online user forum domains: Ubuntu technical help and TripAdvisor topics for New York City \cite{bhatia2010adopting}. 
For each thread, we concatenate the title and initial post to generate passages. For BERT-based models we truncate at 350 wordpiece tokens.  Unlike the BioASQ data, this data generally does not contain specialist knowledge queries. Thus, compared to the collection of question-answer pairs mined from the web, there is less of a domain shift. 

\vspace{-1mm}
\paragraph{NaturalQuestions} Aggregated queries issued to Google Search \cite{NQ} with relevance judgements. 
We convert the original format to a passage retrieval task, where the goal is to retrieval the long answer among all wiki paragraphs \cite{ahmad-etal-2019-reqa}.
We discarded questions whose long answer is either a table or a list.  We evaluate retrieval performance on the  development set as the test set is not publicly available. 
The target collection contains all passages from the development set and is augmented with passages from 2016-12-21 dump of Wikipedia \cite{drqa}. Each passage is also concatenated with title. For BERT-based models passages are truncated at 350 wordpiece tokens. 
This data is different from the previous data in two regards. First, there is a single annotated relevant paragraph per query. This is due to the nature in which the data was curated. Second, this data is entirely ``general domain''.

Dataset statistics are listed in  Appendix~\ref{app:data}.

\subsection{Zero-shot Systems}

\paragraph{BM25} Term-matching systems such as BM25 \cite{robertson1995okapi} are themselves zero-shot, since they require no training resources except the document collection itself. We train a standard BM25 retrieval model on the document collection for each target domain.

\paragraph{ICT} The Inverse Cloze Task (ICT) \cite{lee2019latent} is an unsupervised pre-training objective which randomly masks out a sentence from a passage and creates synthetic sentence-passage pairs representing membership of the sentence in the passage. These masked examples can then used to train or pre-train a retrieval model. \newcite{lee2019latent} showed that masking a sentence with a certain probability, $p$, can both mimic the performance of lexical matching ($p=0$) or semantic matching ($p>0$). 
ICT is \textit{domain-targeted} since training examples are created directly from the relevant  collection. \newcite{chang-ict} showed that ICT-based pre-training outperforms a number of alternatives such as Body First Selection (BFS) or Wiki Link Prediction (WLP) for large-scale retrieval. 

\paragraph{Ngram} \newcite{NVS-van-gysel} proposes to train unsupervised neural retrieval system by extracting ngrams and titles from each document as queries.  Different from ICT, this approach does not mask the extract ngrams from the original document.


\paragraph{QA} The dataset mined from community question-answer forums (Sec.~\ref{sec:qgen}) itself can be used directly to train a neural retrieval model since it comes of the form query and relevant text (passage) pair. This data is naturally occurring and not systematically noisy, which is an advantage. However, the data is not domain-targeted, in that it comes from general knowledge questions. We call models trained on this dataset as QA. Applying a model trained on general domain data to a specific domain with no adaptation is a strong baseline \cite{yilmaz2019cross}.

\paragraph{QGen} The QGen retrieval model trained on the domain-targeted synthetic question-passage pairs described in Section~\ref{sec:qgen}. While this model can contain noise from the generator, it is domain-targeted. 

\paragraph{QGenHyb} This is identical to QGen, but instead of using the pure neural model, we train the hybrid model in Section~\ref{sec:hybrid} setting $\lambda=1.0$ for all models to avoid any domain-targeted tuning. We train the term and neural components independently, combing them only at inference.

All \textbf{ICT}, \textbf{NGram}, \textbf{QA}  and \textbf{QGen} models are trained using the neural architecture from Section~\ref{sec:neuralretrieval}. 
For BioASQ experiments, question and passage encoders are initialized with BioBERT base v-1.1 \cite{biobert}. All other data uses uncased BERT base \cite{bert}.

We can categorize the neural zero-shot models along two dimensions \emph{extractive} vs. \emph{transfer}. ICT and Ngram are extractive, in that they extract exact substrings from a passage to create synthetic questions for model training. Note that extractive models are also unsupervised, since they do not rely on general domain resources. QA is a \emph{direct} cross-domain transfer model, in that we train the model on data from one domain (or general domain) and directly apply it to the target domain for retrieval. QGen models are \emph{in-direct} cross-domain transfer models, in that we use the out-of-domain data to generate resources for model training.

\subsection{Generated Training Datasets}
\label{sec:qgen_data}

\begin{table}[t]
\centering
\resizebox{\columnwidth}{!}{
\begin{tabular}{l | rr r rr}
              & QA & ICT & Ngram & ICT+Ngram & QGen  \\ \hline
BioASQ        & 2.00M & 90.50M & 636.54M & 727.05M & 82.62M \\
NQ            & 2.00M & 71.58M & 356.15M & 427.72M & 84.33M \\
Forum Travel  & 2.00M &  0.30M &   1.25M &   1.54M &  0.26M \\ 
Forum Ubuntu  & 2.00M &  0.42M &   2.07M &   2.49M &  0.43M \\
\hline    
\end{tabular}
}
\vspace{-0.1in}
\caption{Number of (synthetic-question, passage) pairs used in zero-shot experiments.}
\label{tab:stats_zero_shot}
\vspace{-0.1in}
\end{table}

The nature of each zero-shot neural system requires different generated training sets. For ICT, we follow \newcite{lee2019latent} and randomly select at most 5 sentences from a document, with a mask rate of 0.9.
For Ngram models, \newcite{NVS-van-gysel} suggests that retrieval models trained with ngram-order of around 16 was consistently high in quality.  
Thus, in our experiment we also use 16 and move the ngram window with a stride of 8 to allow 8 token overlap between consecutive ngrams.

For QGen models, each passage is truncated to 512 sentence tokens and feed to the question generation system. We also run the question generator on individual sentences from each passage to promote questions that focus on different aspects of the same document. We select at most 5 salient sentences from a passage, where sentence saliency is the max term IDF value in a sentence.

The size of the generated training set for each baseline is shown in Table~\ref{tab:stats_zero_shot}.

\begin{table*}[t]
\centering
\resizebox{0.8\width}{!}{
\begin{tabular}{l |lll |lll |lll |lll}

  \multicolumn{1}{c}{ }& \multicolumn{3}{c}{\textbf{BioASQ 7}} & \multicolumn{3}{c}{\textbf{BioASQ 8}} & \multicolumn{3}{c}{\textbf{Forum Travel}} & \multicolumn{3}{c}{\textbf{Forum Ubuntu}} \\
  \vspace{-0.04in}
  &  & \textbf{Prec} & \textbf{nDCG} &  & \textbf{Prec} & \textbf{nDCG} &  & \textbf{Prec} & \textbf{nDCG} &  & \textbf{Prec} & \textbf{nDCG} \\ 
  & \textbf{MAP} & \textbf{@10} & \textbf{@10} & \textbf{MAP} & \textbf{@10} & \textbf{@10} & \textbf{MAP} & \textbf{@10} & \textbf{@10} & \textbf{MAP} & \textbf{@10} & \textbf{@10} \\

\hline\hline
\multicolumn{13}{c}{\textbf{\textsc{Neural Models}}}\\
\hline\hline

ICT$^\star$ & 9.31$^\ast$ & 3.84$^\ast$ & 11.44$^\ast$ & 9.31$^\ast$ & 3.36$^\ast$ & 11.78$^\ast$ & 3.66$^\ast$ & 11.60$^\ast$ & 12.04$^\ast$ & 8.93$^\ast$ & 21.60$^\ast$ & 23.21$^\ast$ \\

Ngram$^\star$ & 9.17$^\ast$ & 3.86$^\ast$ & 11.53$^\ast$ & 8.81$^\ast$ & 2.84$^\ast$ & 10.74$^\ast$ & 10.00 & 25.60 & 28.53 & 9.44$^\ast$ & 22.00$^\ast$ & 23.90$^\ast$  \\

QA$^\dagger$ & 17.80$^\ast$ & 7.46$^\ast$ & 21.93$^\ast$ & 14.61$^\ast$ & 4.26$^\ast$ & 17.09$^\ast$ & 11.00 & 27.60 & 28.32 & 17.78 & \bs{34.00} & 34.73 \\
\cdashline{1-13}

QGen$^\ddagger$ & \bs{32.45} & \bs{13.48} & \bs{37.23} & \bs{30.32} & \bs{9.36} & \bs{34.53} & \bs{11.79} & \bs{32.00} & \bs{33.34} & \bs{17.97} & 32.40 & \bs{36.11} \\

\hline\hline
\multicolumn{13}{c}{\textbf{\textsc{Term/Hybrid Models}}}\\
\hline\hline

BM25$^\star$ & 45.12$^\ast$ & \bs{20.66} & 50.33$^\ast$ & 38.61$^\ast$ & 11.94$^\ast$ & 42.78$^\ast$ & 15.41$^\ast$ & 37.60 & 39.21 & 16.23$^\ast$ & 31.20$^\ast$ & 35.16$^\ast$ \\

QGenHyb$^\ddagger$ & \bs{46.78} & 20.60 & \bs{52.16} & \bs{41.73} & \bs{12.84} & \bs{46.18} & \bs{18.19} & \bs{40.80} & \bs{43.92} & \bs{21.97} & \bs{39.60} & \bs{43.91} \\
\hline

\hline
\end{tabular}
}
\vspace{-0.05in}
\caption{Zero-shot first-stage retrieval. Unsupervised$^\star$; Out-of-domain$^\dagger$; Synthetic$^\ddagger$. Bold=Best in group.
Statistically significant differences (permutation test, $p < 0.05$) from the last row of each group are marked by $^\ast$.
}
\label{ParagraphRetrieval}
\vspace{-0.1in}
\end{table*}

\section{Results and Discussion}

Our main results are shown in Table~\ref{ParagraphRetrieval}.
We compute Mean Average Precision over the first N\footnote{BioASQ: N=100; and Forum: N=1000.} results (MAP), Precision@10 and nDCG@10 \cite{manning2008introduction} with TREC evaluation script\footnote{\url{https://trec.nist.gov/trec_eval/}}. All numbers are in percentage.

Accuracy of pure neural models are shown in the upper group of Table~\ref{ParagraphRetrieval}.
First, we see that both QA and QGen consistently outperform neural baselines such as ICT and Ngram that are based on sub-string masking or matching.
Matching on sub-strings likely biases the model towards memorization instead of learning salient concepts of the passage. 
Furthermore, query encoders trained on sub-strings are not exposed to many questions, which leads to adaptation issues when applied to true retrieval tasks.
Comparing QGen with QA, typically QGen performs better, especially for specialized target domains. This suggests that domain-targeted query generation is more effective for domain shift than
direct cross-domain transfer \cite{yilmaz2019cross}.


Performance of term-based models and hybrid models are shown in Table~\ref{ParagraphRetrieval} (bottom).  
We can see that BM25 is a very strong baseline. 
However, this could be an artifact of the datasets as the queries are created by annotators who already have the relevant passage in mind.  
Queries created this way typically have large lexical overlapping with the passage, thus favoring term matching based approaches like BM25.  
This phenomenon has been observed by previous work \newcite{lee2019latent}.
Nonetheless, the hybrid model outperforms BM25 on all domains, and the improvements are statistically significant on 9/12 metrics.
This illustrate that term-based model and neural-based model return complementary results, 
and the proposed hybrid approach effectively combines their strengths. 


For NaturalQuestions since there is a single relevant passage annotation, we report Precision@1 and Mean reciprocal rank (MRR)\footnote{MRR = MAP when there is one relevant item.}.
Results are show in Table~\ref{NGParagraphRetrieval}. 
We can see here that while QGen still significantly outperform other baselines, the gap between QGen and QA is smaller.
Unlike BioASQ and Forum datasets, NaturalQuestions contains general domain queries,
which aligns well with the question-answer pairs for training the QA model.
Another difference is that NaturalQuestions consists of real information seeking queries,
in this case QGen performs better than BM25. 


\begin{table}[t]
\centering
\resizebox{0.65\columnwidth}{!}{
\begin{tabular}{l |ll}
  & \textbf{MRR} & \textbf{Prec@1} \\ \hline
BM25$^\star$ & 6.63$^\ast$ & 1.84$^\ast$\\ 
ICT$^\star$ & 4.62$^\ast$ & 1.58$^\ast$\\ 
Ngram$^\star$ & 7.22$^\ast$ & 3.05$^\ast$\\ 
QA$^\dagger$ & 11.14$^\ast$ & 4.35$^\ast$\\ 
\cdashline{1-3}
QGen$^\ddagger$ & \bnh{14.93} & \bs{\bnh{6.21}} \\ 
QGenHyb$^\ddagger$ & \bs{16.73} & 6.05 \\ 
\hline
\textit{Supervised} & \textit{33.68} & \textit{17.33} \\ 
\hline
\end{tabular}
}
\vspace{-0.05in}
\caption{Zero-shot ad-hoc retrieval for Natural Questions. Unsupervised$^\star$; Out-of-domain$^\dagger$; Synthetic$^\ddagger$. Bold=Best; Underline=Best non-hybrid.
Baselines with statistically significant differences (permutation test, $p < 0.05$) from QGen are marked by $^\ast$.
}
\label{NGParagraphRetrieval} 
\vspace{-0.1in}
\end{table}

\subsection{Zero-shot vs.\ Supervised}
\label{sec:zs_v_sup}

One question we can ask is how close to the state-of-the-art in supervised passage retrieval are these zero-shot models. To test this we looked at BioASQ 8 dataset and compare to the top-participant systems.\footnote{\texttt{participants-area.bioasq.org}} Since BioASQ provides annotated training data, the top teams typically use supervised models with a first-stage retrieval plus rescorer architecture. For instance, the AUEB group, which is the top or near top system for BioASQ 6, 7 and 8, uses a BM25 first-stage retrieval model plus a supervised neural rescorer \cite{brokos2018aueb,pappas2019aueb}.

In order to make our results comparable to participant systems, we return only 10 passages per question (as per shared-task guidelines) and use the official BioASQ 8 evaluation software.

Table~\ref{tab:zs_v_sota} shows the results for three zero-shot systems (BM25, QGen and QGenHyb) relative to the top 4 systems on average across all 5 batches of the shared task. We can see the QGenHyb performs quite favorably and on average is indistinguishable from the top systems. This is very promising and suggests that top-performance for zero-shot retrieval models is possible. 

A natural question is whether improved first-stage model plus supervised rescoring is additive. The last two lines of the table takes the two-best first-stage retrieval models and adds a simple BERT-based cross-attention rescorer \cite{Nogueira2019PassageRW, macavaney2019cedr}. We can see that, on average, this does improve quality. Furthermore, having a better first-stage retriever (QGenHyb vs. BM25) makes a difference.

As noted earlier, on BioASQ, BM25 is a very strong baseline. This makes the BM25/QGenHyb zero-shot models highly likely to be competitive. When we look at NaturalQuestions, where BM25 is significantly worse than neural models, we see that the gap between zero-shot and supervised widens substantially. The last row of Table~\ref{NGParagraphRetrieval} shows a model trained on the NaturalQuestions training data, which is nearly 2-3 times more accurate than the best zero-shot models.
Thus, while zero-shot neural models have the potential to be competitive with supervised counterparts, the experiments here show this is data dependant.

\subsection{Learning Curves}

\begin{figure}[t!]
  \center{
    \begin{tikzpicture}
      \begin{axis}[
        width=2.5in,
        height=2.5in,
        title={},
        ylabel={},
        xmin=-1, xmax=101,
        ymin=5, ymax=40,
        xtick={0, 20, 40, 60, 80, 100},
        ytick={10, 20, 30, 40},
        yticklabels={10\%, 20\%, 30\%, 40\%},
        xticklabels={1\%, 5\%, 10\%, 20\%, 40\%, 100\%},
        tick label style={font=\scriptsize},
        legend pos=outer north east,
        xmajorgrids=true,
        ymajorgrids=true,
        grid style=dashed,
        legend style={font=\scriptsize},
        ]
        \addplot [color=blue,mark=square] table[x=TrainingData,y=P@10]{\learncurve};
        \addplot [color=red,mark=diamond] table[x=TrainingData,y=NDCG@10]{\learncurve};
        \addplot [color=green,mark=*] table[x=TrainingData,y=mAP]{\learncurve};
        \legend{P@10, NDCG@10, mAP}
      \end{axis}
    \end{tikzpicture}
  }
  \vspace{-0.05in}
  \caption{MAP on BioASQ7 (y-axis) w.r.t. the \% of documents used for synthesizing queries (x-axis).}
  \label{fig:coverage}
\end{figure}
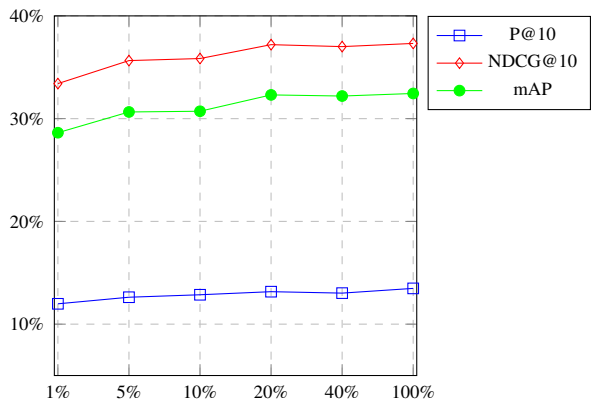

Since our approach allows us to generate queries on every passage of the target corpus, one question is that whether retrieval system trained this way simply memorizes the target corpus or it also generalize on unseen passages. Furthermore, from an efficiency standpoint, how many synthetic training examples are required to achieve maximum performance.
To answer these questions, we uniformly sample a subset of documents and then generate synthetic queries only on that subset.  
Results on BIOASQ 7 are shown in Figure~\ref{fig:coverage}, where x-axis denotes the percentage of sampled documents.
We can see that retrieval accuracy improves as passage coverage increases. 
The peak is achieved when using a $20\%$ subset, which covers $21\%$ of the reference passages.
This is not surprising because the number of frequently discussed entities/topics are typically limited, and a subset of the passages covers most of them.   
This result also indicates that the learned system does generalize, otherwise optimal performance would be seen with $100\%$ of the data.

\begin{table}[t]
\centering
\resizebox{\columnwidth}{!}{
\begin{tabular}{l | ccccc|c}
 & B1 & B2 & B3 & B4 & B5 & Avg. \\ \hline
BM25 & 31.7 & 27.8 & 40.4 & 40.1 & 41.8 & 36.3 \\
QGen & 28.9 & 20.3 & 30.7 & 29.0 & 33.1 & 28.4 \\
QGenHyb & 34.8 & 31.3 & 43.4 & 41.9 & 45.3 & 39.3 \\ \cdashline{1-7}
AUEB-1 & 33.6 & 31.8 & \bs{44.4} & 40.1 & 46.0 & 39.2 \\
pa & 33.5 & \bs{33.0} & 43.5 & 36.0 & \bs{48.3} & 38.9 \\
bioinfo-3 & 34.0 &	31.7 & 43.7 & 40.2 & 46.7 & 39.2 \\
DeepR-test & 30.7 & 29.1 & 43.5 & 39.8 & 47.5 & 38.1 \\ \hline
BM25$\rightarrow$resc. & 33.9 & 29.2 & 42.4 & 42.5 & 457 & 38.7\\
QGenHyb$\rightarrow$resc. & \bs{37.5} & 31.2 & 43.0 & \bs{43.6} & 46.6 & \bs{40.4} \\
\hline    
\end{tabular}
}
\vspace{-0.05in}
\caption{MAP for zero-shot models (above dashed lined) vs.\ supervised models (below dashed line) on BioASQ8 document retrieval. B1-B5 is batch 1-5.}
\label{tab:zs_v_sota}
\vspace{-0.1in}
\end{table}

\subsection{Generation vs.\ Retrieval Quality}

Another interesting question is how important is the quality of the question generator relative to retrieval performance.
Below we measured generation quality (via Rouge-based metrics \cite{lin2002manual}) versus retrieval quality for three systems. The base generator contains 12 transformer layers, the lite version only uses the first 3 layer.  The large one contains 24 transformer layers and each layer with larger hidden layer size, 4096, and more attention heads, 16.   
Retrieval quality was measured on BIOASQ 7 and generation quality with a held out set of the community question-answer data set.
Results are shown in Table~\ref{tab:gen_v_ir}. We can see that larger generation models lead to improved generators. However, there is little difference in retrieval metrics, 
suggesting that large domain targeted data is the more important criteria.

\section{Conclusion}
We study methods for neural zero-shot passage retrieval and find that domain targeted synthetic question generation coupled with hybrid term-neural first-stage retrieval models consistently outperforms alternatives. Furthermore, for at least one domain, approaches supervised quality. While out of the scope of this study, future work includes further testing the efficacy of these first-stage models in a full end-to-end system (evaluated briefly in Section~\ref{sec:zs_v_sup}), as well as for pre-training supervised models \cite{chang-ict}.

\pgfplotstableread[row sep=\\,col sep=&]{
  TrainingData & mAP & P@10 & NDCG@10  \\ 
    0 & 28.63 & 11.98 & 33.41 \\
    20 & 30.65 & 12.62 & 35.65 \\
    40 & 30.72 & 12.86 & 35.85 \\
    60 & 32.31 & 13.16 & 37.2 \\
    80 & 32.2 & 13.02 & 37.01 \\
  100 & 32.45 & 13.48 & 37.32 \\
}\learncurve

\begin{table}[t]
\centering
\resizebox{0.9\columnwidth}{!}{
\begin{tabular}{l | cc|ccc}
 \multicolumn{1}{c}{} & \multicolumn{2}{c}{\textbf{Generation}} & \multicolumn{3}{c}{\textbf{Retrieval}} \\
 \vspace{-0.04in}
& \textbf{Rouge} & \textbf{Rouge} &  & \textbf{Prec} & \textbf{nDCG} \\
& \textbf{1} & \textbf{L} & \textbf{MAP} & \textbf{@10} & \textbf{@10} \\ \hline
Lite & 23.55 & 21.90  & 32.50 & \textbf{13.48} & 37.23 \\
Base & 26.20 & 24.23 & \textbf{32.86} & 13.42 & \textbf{37.96} \\
Large & \textbf{26.81} & \textbf{24.90} & 32.61 & 13.34 & 37.53 \\
\hline    
\end{tabular}
}
\vspace{-0.05in}
\caption{Generation quality vs.\ retrieval metrics.}
\label{tab:gen_v_ir}
\vspace{-0.1in}
\end{table}

\section*{Acknowledgements}

We thank members of Google Research Language for feedback on this work. In particular, Gon\c{c}alo Sim\~{o}es gave detailed feedback on an early draft of this work, and Shashi Narayan evaluated the question generation quality.

\bibliography{anthology,eacl2021}
\bibliographystyle{acl_natbib}

\input{appendix-content}

\end{document}

%% file: appendix-content.tex
\appendix

\section{Data}
\label{app:data}

Statistics on each evaluation set are listed in Table~\ref{tab:statis_supervised}.
Document collection of ``BioASQ'' comes from MEDLINE articles, and we remove roughly 10M articles that only contains a title. 
Furthermore, for BioASQ 7B and BioASQ 8B we only keep articles published before 2018 December 31 and 2019 December 31, respectively.
On ``Forum'', we remove threads with empty posts.
On "NQ" since there is at most one passage annotated as relevant for each question, and we also remove questions that have no answer, thus the number of questions equal to the number of reference passages.
Besides zero-shot experiments, we also conduct supervised experiments on NQ, where we randomly samples $5\%$ question from the training data as development set.  
This yields a training and development set with 70,393 and 3,704 (question, passage) pairs, respectively.

The data resources can be downloaded from the following websites
\begin{itemize}
    \item BioASQ: \url{http://participants-area.bioasq.org/}
    \item Forum: \url{http://sumitbhatia.net/source/datasets.html}
    \item Natural Questions:  \url{https://github.com/google/retrieval-qa-eval}
    \item Pubmed / Medline:  \url{https://www.nlm.nih.gov/databases/download/pubmed_medline.html}
    \item Stackexchange: \url{http://archive.org/details/stackexchange}
    \item Yahoo! Answers: \url{http://webscope.sandbox.yahoo.com/catalog.php?datatype=l}
    \item BioBERT: \url{https://github.com/dmis-lab/biobert}
    \item BERT: \url{https://github.com/google-research/bert}
\end{itemize}
To the extent that we pre-process the data, we will release relevant tools and data upon publication.

\begin{table}[t]
\centering
\resizebox{\columnwidth}{!}{
\begin{tabular}{l | ccccc}
\vspace{-0.04in}
           & BioASQ7 & BioASQ8 & NQ & ForumTravel & ForumUbuntu \\ \hline
Q  & 500 & 500 & 1772  & 25     & 25     \\
R  & 2349 & 1646 & 1772  & 1,538  & 1,188 \\
C  & 50M & 53.4M & 29.5M & 82,669 & 106,642 \\
\hline    
\end{tabular}
}
\vspace{-0.1in}
\caption{Statistics on each evaluation set. ``Q" denotes the number of unique questions. ``R" denotes the total number of annotated reference passages. ``C" is the number of passages in the target collection.}
\label{tab:statis_supervised}
\end{table}


\section{Question Generation Details}
\label{app:qgen_details}
Our question generation follows the same implementation of \citet{berts2s}.  
Both the encoder and decoder share the same network structure. 
Parameter weights are also shared and are initialized from a pretrained RoBERTa \cite{roberta} checkpoints.
Training data is processed with sentencepiece \cite{sentencepiece} tokenization.
We truncate answers to 512 sentencepiece tokens, and limit decoding to at most 64 steps.
The training objective is the standard cross entropy.
We use Adam \cite{adam} with learning rate of 0.05, $\beta_1 = 0.9$, $\beta_2 = 0.997$ and $\epsilon = 1e-9$.
Learning rate warmup over the first 40,000 steps.
Training batch size for the ``lite", ``base" and ``large" models are 256, 128 and 32 respectively.
All models are trained on a ``4x4" slice of v3 Google Cloud TPU.
At inference, results from using beam search decoding usually fall short of diversity, 
thus we use greedy decoding to speed up question generation.

\section{Neural Model Details}

\subsection{Zero shot Retrieval Models}
\subsubsection{Development Set}

Since we are investigating zero-shot scenario where there is no annotated development set available,
hyperparameters are set by following best practice reported in previous work. We thus do not have development set numbers. However, in the hyperparameters section below, we do use a subset of the zero-shot training data to test training convergence under different parameters. 

\subsubsection{Data Generation}

For ICT task, we follow \newcite{lee2019latent} and randomly select at most 5 sentences from a document, with a mask rate of 0.9.
For Ngram models, \newcite{NVS-van-gysel} suggests that retrieval models trained with N larger than 16 consistently outperform those trained
with N smaller than 8.  
In addition, further increase N from 16 has little effect on retrieval accuracy.  
Thus, in our experiment we set N to 16 and move the ngram window with a stride of 8 to allow 8 token overlap between consecutive ngrams.
For QGen models, each passage is truncated to 512 sentence tokens and feed to the question generation system.
Besides, we also run question generator on individual sentences from each document to promote questions that focus on different aspects of the same document.  
In particular, we select at most top 5 salient sentences from a document, where salience of a sentence is measure as the max IDF value of terms in that sentence. 
We then feed these sentences to the question generator.

\subsubsection{Hyperparameters}

For zero-shot neural retrieval model training, we uniformly sample of a subset of 5K (question, document) pairs from the training data as a noisy development set.
Instead of finding the best hyperparameter values, 
we use this subset to find the largest batch size and learning rate that lead the training to converge \cite{smith-batch-size}.
Take batch size for example, we always start from the largest batch that can fit in the memory of a ``8x8'' TPU slice.
We gradually decrease the batch size by a factor of 2 if the current value causes training diverge.
More details of hyperparameter values of each task are listed in Table~\ref{tab:de-hyperparameter}.
Note on Forum data, the maximum batch size for QGen is much larger than other tasks.  
Looking into the data, we found that queries generated by ICT or Ngram task on Forum data tends to contain higher percentage of noisy sentences or ngrams that are either ill-relevant to the topic or too general.  
For example, ``suggestions are welcomed", ``any ideas for things to do or place to stay".
We train each model for 10 epochs, but also truncate training steps to 200,000 to make training time tractable.

For BM25, the only two hyperparameters are $k$ and $b$. We set these to $k=1.2$ and $b=0.75$ as advised by \newcite{manning2008introduction}.

For the hybrid model QGenHyb, the only hyperparameter is $\lambda$. We set this to 1.0 without any tuning, since this represented an equal trade-off between the two models and we wanted to keep the systems zero-shot. However, we did try experimentations. For BioASQ 8b and Forum Ubuntu, values near 1.0 were actually optimal. For BioASQ 7b and Forum Travel, values of 2.0 and 2.1 were optimal and led to improvements in MAP from $0.468\rightarrow0.474$ and $0.181\rightarrow0.188$, respectively.

\subsection{Supervised Models}

We also train supervised models on BioASQ and NQ, where we use the development set to do early stopping. For BioASQ, our developement set is data from BioASQ 5 (i.e., disjoint from BioASQ 7 and 8). The development set MAP of our supervised model reranking a BM25 system on this data is 52.1, compared to the BioASQ 8 scores of 38.7. For NQ, the MRR on the development set is 0.141.
All other hyperparameters remain the same except we use a smaller batch size of 1024, as we observe that using large batch causes the model quickly overfit the training data. 
This may due to the number of training examples is 2 orders of magnitude smaller compared to zero-shot setting. For our BioASQ supervised model we follow \newcite{pappas2019aueb} and train it with binary cross-entropy using the top 100 BM25 results as negatives.

\begin{table}[t]
\centering
\resizebox{\columnwidth}{!}{
\begin{tabular}{c l |cc}

 \vspace{-0.04in}
  & & \textbf{Learning Rate} & \textbf{Batch Size}  \\ \hline

\multirow{4}{*}{\rotatebox{90}{\footnotesize\textbf{BioASQ}}} & ICT       & 1e-5 & 8192  \\
& Ngram       & 1e-5 & 8192  \\
& QGen        & 1e-5 & 8192  \\
\hline

\multirow{4}{*}{\rotatebox{90}{\makecell{\footnotesize \textbf{Forum}\\\textbf{Travel}}}} & ICT & 2e-6 & 1024  \\
& Ngram & 2e-6 & 1024  \\
& QGen & 2e-6 & 4096  \\
\hline

\multirow{4}{*}{\rotatebox{90}{\makecell{\footnotesize \textbf{Forum}\\\textbf{Ubuntu}}}} & ICT & 1e-6 & 512  \\
& Ngram & 1e-6 & 512  \\
& QGen & 1e-6 & 4096   \\
\hline

\multirow{4}{*}{\rotatebox{90}{\footnotesize\textbf{NQ}}} & ICT       & 1e-5 & 6144  \\
& Ngram       & 1e-5 & 6144  \\
& QGen        & 1e-5 & 6144  \\
\hline

\end{tabular}
}
\caption{Hyperparameters}
\label{tab:de-hyperparameter} 
\end{table}








\subsection{Computational Resources}

\subsubsection{Question Generation}

To train the question generator on 2M questions,

\begin{itemize}
    \item We used a ``4x4" slice of v3 Google Cloud TPU.
    \item Training time ranges from 20 hours for the lite model and 6 days for the large model. 
\end{itemize}

Once trained, we need to run the generator over our passage collection. 

\begin{itemize}
    \item We distributed computation and used 10,000 machines (CPUs) over the collection.
    \item For BioASQ, the largest dataset, it took less than 40 hours to generate synthetic questions.
\end{itemize}

We initialize question generation models from either RoBerta base or Roberta large checkpoint \cite{roberta}, and the total number of trainable parameters are 67M for the lite model, 152M for the base model and 455M for the large model.

\subsubsection{Neural Retrieval Model}

To train the retrieval models, we need to train the query and passage encoders. 
We share parameters between the two encoders and initialize them from either base BERT \cite{bert} or BioBERT \cite{biobert} checkpoint.  
Thus retrieval models trained on BioASQ have 108M trainable parameters and retrieval models trained on NQ and Forum data have 110M trainable parameters.
After training, we need to run the passage encoder over every passage in the collection to create the nearest neighbour backend.

\begin{itemize}
    \item Depending on the training batch size, we use either an ``8x8'' or ``4x4'' TPU slice.
    \item Training the "ngram" model on BioASQ took the longest time, which completes in roughly 30 hours.
    \item Indexing BioASQ, which is our largest passage collection, with 4000 CPUs which took roughly 4 hours.
\end{itemize}

Having trained models, the inference task is to encode a query with the neural model and query the distributed nearest neighbour backend to get the top ranked passages. The relevant resources are:

\begin{itemize}
    \item We encode queries on a single CPU.
    \item Our distributed nearest neighbour search uses 20 CPUs to serve the collections.
    \item For BioASQ, our largest collection, to run the inference on the test sets of 500 queries took roughly 1m57s. This is approximately 0.2s per instance to encode the query, run brute-force nearest neighbour search on 10s of millions of examples and return the result.
\end{itemize}